\documentclass[12pt]{article}

\usepackage{graphicx}
\usepackage{amstext}
\usepackage{amsfonts}

\def\ltwid{\mathrel{\raise.3ex\hbox{$<$\kern-.75em\lower1ex\hbox{$\sim$}}}}
\def\gtwid{\mathrel{\raise.3ex\hbox{$>$\kern-.75em\lower1ex\hbox{$\sim$}}}}

\def\square{\kern1pt\vbox{\hrule height 1.2pt\hbox{\vrule width 1.2pt\hskip 3pt
   \vbox{\vskip 6pt}\hskip 3pt\vrule width 0.6pt}\hrule height 0.6pt}\kern1pt}
\def\overleftrightarrow#1{\vbox{\ialign{##\crcr
     $\leftrightarrow$\crcr\noalign{\kern-1pt\nointerlineskip}
     $\hfil\displaystyle{#1}\hfil$\crcr}}}

\begin{document}

\begin{titlepage}

\begin{flushright}
UFIFT-QG-13-04 \\
CCTP-2013-10
\end{flushright}

\vskip 1cm

\begin{center}
{\bf Hartree Approximation to the One Loop Quantum Gravitational
Correction to the Graviton Mode Function on de Sitter}
\end{center}

\vskip .5cm

\begin{center}
P. J. Mora$^{1a}$, N. C. Tsamis$^{2b}$ and R. P. Woodard$^{1c}$
\end{center}

\begin{center}
\it{$^{1}$ Department of Physics, University of Florida \\
Gainesville, FL 32611, UNITED STATES}
\end{center}

\begin{center}
\it{$^{2}$ Institute of Theoretical Physics \& Computational Physics \\
Department of Physics University of Crete \\
GR-710 03 Heraklion, HELLAS}
\end{center}

\vskip .5cm

\begin{center}
ABSTRACT
\end{center}
We use the Hartree approximation to the Einstein equation on de
Sitter background to solve for the one loop correction to the
graviton mode function. This should give a reasonable approximation
to how the ensemble of inflationary gravitons affects a single
external graviton. At late times we find that the one loop
correction to the plane wave mode function $u(\eta,k)$ goes like $G
H^2 \ln(a)/a^2$, where $a$ is the inflationary scale factor. One
consequence is that the one loop corrections to the ``electric''
components of the linearized Weyl tensor grow compared to the tree
order result.

\begin{flushleft}
\end{flushleft}

\vskip .5cm

\begin{flushleft}
$^{a}$ e-mail: pmora@phys.ufl.edu \\
$^{b}$ e-mail: tsamis@physics.uoc.gr \\
$^{c}$ e-mail: woodard@phys.ufl.edu
\end{flushleft}

\end{titlepage}

\section{Introduction}\label{Intro}

Primordial cosmological perturbations are believed originate in the
scalars and gravitons produced by inflation \cite{perts}. These
inflationary scalars and gravitons also interact among themselves,
and they should affect the dynamics of other particles. One analyzes
the latter sort of effect by first computing the contribution of
inflationary scalars and/or gravitons to the appropriate 1PI
(one-particle-irreducible) 2-point function, then using this 1PI
2-point function to quantum-correct the linearized effective field
equation of the particle whose dynamics are being studied.

The past decade has witnessed a number of studies of this type. It
is easier to work with massless, minimally coupled (MMC) scalars
than with gravitons, so scalar effects were probed first:
\begin{itemize}
\item{When a MMC scalar is endowed with a quartic self-interaction
the scalar mode function behaves as if its mass were growing
\cite{BOW};}
\item{Yukawa-coupled MMC scalars cause fermions to develop a growing
mass, even when the scalar has zero potential \cite{Yukawa};}
\item{Charged MMC scalars induce so much vacuum polarization that the
photon develops mass \cite{SQED} and there are corresponding
corrections to electromagnetic forces \cite{DW} and}
\item{Gravitationally coupled MMC scalars do not induce any secular
change in the graviton mode function \cite{SPW}.}
\end{itemize}
Three studies have been made of what inflationary gravitons do to
other particles:
\begin{itemize}
\item{There is a slow secular growth in the field strength of
massless fermions \cite{MW1,enh}, driven by the spin-spin interaction
\cite{MW2}, and a much larger effect driven by a small, nonzero mass
\cite{SPM};}
\item{The absence of any spin-spin coupling prevents inflationary
gravitons from having a comparable effect on MMC scalars \cite{KW};
and}
\item{Inflationary gravitons induce a slow secular growth in the
electric components of the photon field strength but no comparable
growth in the magnetic field \cite{LW}.}
\end{itemize}

The purpose of this paper is to begin the study of what inflationary
gravitons do to other gravitons. The first step of an exact analysis
would be to compute the graviton self-energy at one loop order in de
Sitter background. That has been done \cite{TW1}, but only using an
old formalism for which dimensional regularization cannot be
employed, so the result is only valid away from coincidence. The
formalism required for a fully dimensionally regulated computation
has since then been developed \cite{TW2}, but its application to the
graviton self-energy has not yet been completed. In the meantime we
can gain a qualitative understanding of the potential results by
employing the Hartree approximation \cite{Howie}. This has been
shown to predict the correct time dependence for the effects of MMC
scalars on photons \cite{DDPT,PW}, the effects of gravitons on
fermions \cite{MW1,enh}, and for the effects of gravitons on photons
\cite{LW}.

This paper contains five sections, of which the first is this
Introduction. In section \ref{Feynman} we describe the de Sitter
background geometry, the quantum gravity Lagrangian whose field
equations we will solve in the Hartree approximation and our choice
of gauge. The Hartree approximation to the linearized effective
field equation is derived in section \ref{hartree}, and solved for
plane wave gravitons in section \ref{Deltau}. Our discussion
comprises section \ref{Epilogue}.

\section{Feynman Rules}\label{Feynman}

In this section we give the Feynman rules that we will use in this
study. These will derive from the Lagrangian of quantum gravity
whose dynamical field corresponds to a conformally rescaled graviton
field. We will employ a generalized version of the de Donder gauge
fixing term and we will present the associated ghost Lagrangian. All
this will be done in the context of conventional perturbation theory
around a de Sitter background, so we find it necessary to start this
section with a review of de Sitter space.

We are interested in the effects of gravitons produced during
primordial inflation. Cosmological observations \cite{WMAP5,Yun}
support the idea that de Sitter space can be considered as a
paradigm for inflation. However, because we are employing de Sitter
as an approximation for the true inflationary background --- which
is a homogeneous, isotropic and spatially flat geometry --- we want
to work on the open coordinate submanifold of the full de Sitter
geometry. Because we shall be using dimensional regularization we
work in $D$-dimensional conformal coordinates $x^\mu=(\eta,x^i)$,
where
\begin{equation}
-\infty < \eta<0 \;\;\;\; , \;\;\;\; -\infty <x^i < \infty \;.
\end{equation}
We express the full metric as,
\begin{equation}
\label{metric}
g_{\mu\nu} = a^2[\eta_{\mu\nu}+\kappa h_{\mu\nu}] \equiv
a^2\widetilde{g}_{\mu\nu}\;.
\end{equation}
Here $a(\eta) \equiv -1/H\eta$ is the scale factor, $H$ is the
(constant) Hubble parameter of de Sitter, $\eta_{\mu\nu}$ is the
Lorentz metric with spacelike signature, $h_{\mu\nu}$ is a
perturbation to this background which we identify with the graviton
field (whose indices are raised and lowered with $\eta_{\mu\nu}$),
and $\kappa^2 \equiv 16 \pi G$ is the loop counting parameter of
quantum gravity. We note that $\widetilde{g}^{\alpha\beta}$ inverts
its covariant counterpart,
\begin{equation}
\widetilde{g}^{\alpha\beta} = \eta^{\alpha\beta} - \kappa
h^{\alpha\beta} + \kappa^2 h^{\alpha\rho} h_\rho^\beta - ...
\end{equation}

The Einstein-Hilbert Lagrangian for quantum gravity is,
\begin{equation}
\label{lag} {\cal L}_{\text{inv}} = \kappa^{-2} \sqrt{-g} \, \Bigl[R
- (D\!-\!2) \Lambda \Bigr] \; ,
\end{equation}
where $R$ is the $D$-dimensional Ricci scalar and $\Lambda \equiv
(D-1) H^2$ is the cosmological constant. Expanding this Lagrangian
with the metric (\ref{metric}) and extracting a presumably
irrelevant surface term (whose expression we will not write here),
we obtain \cite{TW2}:
\begin{eqnarray}
\label{Linv} \lefteqn{{\cal L}_{\text{inv}} - {{\cal S}^\mu}_{,\mu}
= \Bigl(\frac{D}{2} \!-\!1 \Bigr) H a^{D-1} \sqrt{-\widetilde{g}} \,
\widetilde{g}^{\rho\sigma} \widetilde{g}^{\mu\nu} h_{\rho\sigma,\mu}
h_{\nu0} + a^{D-2} \sqrt{-\widetilde{g}} \,
\widetilde{g}^{\alpha\beta} \widetilde{g}^{\rho\sigma}
\widetilde{g}^{\mu\nu} } \nonumber \\
& & \hspace{1.8cm} \times \left(\frac12 h_{\alpha\rho,\mu}
h_{\nu\sigma,\beta} \!-\! \frac12 h_{\alpha\beta,\rho}
h_{\sigma\mu,\nu} \!+\! \frac14 h_{\alpha\beta,\rho}
h_{\mu\nu,\sigma} \!-\! \frac14 h_{\alpha\rho,\mu}
h_{\beta\sigma,\nu} \right) . \qquad
\end{eqnarray}
We fix the gauge by adding an analogue of the de Donder term used in
flat space \cite{TW2},
\begin{equation}
\label{gauge} {\cal L}_{\text{gf}} = - \frac1{2} a^{D-2}
\eta^{\mu\nu} F_\mu F_\nu \;\;,\;\; F_\mu = \eta^{\rho\sigma} \bigg[
h_{\mu\rho,\sigma} - \frac1{2} h_{\rho\sigma,\mu} + (D \!-\!2) H a
h_{\mu\rho} \delta^0_{\sigma} \bigg] \;.
\end{equation}
Because space and time components are treated differently it will
convenient to define the purely spatial parts of the Minkowski
metric and the Kronecker delta function,
\begin{equation}
\overline{\eta}_{\mu\nu} \equiv \eta_{\mu\nu} \!+\! \delta^0_{\mu}
\delta^0_{\nu} \;\;\; , \;\;\; \overline\delta^\mu_\nu \equiv
\delta^\mu_\nu \!-\! \delta^\mu_0\delta^0_\nu \;\;\; , \;\;\;
\overline{\eta}^{\mu\nu} \equiv \eta^{\mu\nu} \!+\! \delta^{\mu}_{0}
\delta^{\nu}_{0} \; .
\end{equation}
The graviton kinetic operator can be found by partially integrating
the quadratic part ${\cal L}_{\text{inv}}+{\cal L}_{\text{gf}}$ to obtain
the form $\frac1{2} h^{\mu\nu} {{\cal D}_{\mu\nu}}^{\rho\sigma}
h_{\rho\sigma}$, where
\begin{eqnarray}
\label{kinop} \lefteqn{{{\cal D}_{\mu\nu}}^{\rho\sigma} =
\left[\frac{1}{2} \overline{\delta}^{(\rho}_\mu
\overline{\delta}^{\sigma)}_{\nu} \!-\! \frac{1}{4} \eta_{\mu\nu}
\eta^{\rho\sigma} \!-\! \frac{1}{2(D-3)} \, \delta_\mu^0
\delta^0_\nu \delta^\rho_0 \delta^\sigma_0 \right] {\cal D}_A }
\nonumber \\
& & \hspace{5cm} + \delta^{0}_{(\mu}
\overline{\delta}^{(\rho}_{\nu)} \delta^{\sigma)}_0 {\cal D}_B +
\frac1{2} \bigg(\frac{D\!-\!2}{D\!-\!3} \bigg) \delta^0_\mu
\delta^0_\nu \delta^\rho_0 \delta^\sigma_0 {\cal D}_C \; . \qquad
\end{eqnarray}
The three scalar differential operators are,
\begin{equation}
{\cal D}_A \equiv \partial_\mu\left(a^{D-2}\eta^{\mu\nu}\partial_\nu
\right) \; , \; {\cal D}_B \equiv {\cal D}_A - (D\!-\! 2) a^D H^2 \;
, \; {\cal D}_C \equiv {\cal D}_A - 2(D\!-\!3) a^D H^2 \;.
\label{DA}
\end{equation}
The associated ghost Lagrangian is,
\begin{eqnarray}
\lefteqn{{\cal L}_{\text{ghost}} \equiv  -a^{D-2}
\overline\omega^\mu \delta F_\mu } \\
& & \hspace{-.5cm} = \overline{\omega}^\mu
\left(\overline{\delta}^\nu_\mu {\cal D}_A \!+\! \delta^0_\mu
\delta^\nu_0 {\cal D}_B\right) \omega_\nu \!-\! 2 \kappa a^{D-2}
\overline{\omega}^{\mu,\nu} \left(h^\rho_{(\mu}
\partial_{\nu)} \!+\! \frac{1}{2} {h_{\mu\nu}}^{,\rho} \!-\! H a
h_{\mu\nu} \delta^\rho_0 \right) \omega_\rho \nonumber \\
& & \hspace{4cm} + \kappa \left( a^{D-2} \overline{\omega}^\mu
\right)_{,\mu} \left(h^{\rho\sigma}
\partial_\sigma \!+\! \frac{1}{2} h_\sigma^{\sigma,\rho} \!-\! H a h
\delta^{\rho}_0 \right) \omega_\rho \; . \qquad
\end{eqnarray}
The ghost and graviton propagator in this gauge can be written in a
simple form as a sum of constant tensor factors times scalar propagators,
\begin{eqnarray}
\label{ghostProp}
i[_{\mu}\Delta_{\nu}](x;x') & = & \overline{\eta}_{\mu\nu} \,
i\Delta_A(x;x') - \delta^0_\mu \delta^0_\nu \, i\Delta_B(x;x') \;, \\
\label{gravProp} i[_{\mu\nu}\Delta_{\alpha\beta}](x;x') & = &
\displaystyle\sum_{I=A,B,C}[_{\mu\nu}T^I_{\alpha\beta}] \,
i\Delta_I(x;x').
\end{eqnarray}
The tensor factors are given by,
\begin{eqnarray}
\label{TensFact} [\mbox{}_{\mu\nu}T^A_{\alpha\beta}] =
2\overline{\eta}_{\mu(\alpha} \overline{\eta}_{\beta)\nu} -
\frac{2}{D\!-\!3} \, \overline{\eta}_{\mu\nu}
\overline{\eta}_{\alpha\beta} \quad ,\quad {[_{\mu\nu}}
T^B_{\alpha\beta}] = -4\delta^0_{(\mu}\overline{\eta}_{\nu)(\alpha}
\delta^0_{\beta)} \; , \nonumber \\
{[_{\mu\nu}}T^C_{\alpha\beta}] = \frac{2}{(D\!-\!2)(D\!-\!3)}
\Bigl[\overline{\eta}_{\mu\nu} + (D\!-\!3 ) \delta^0_\mu
\delta^0_\nu\Bigr] \Bigl[\overline{\eta}_{\alpha\beta} + (D\!-\!3)
\delta^0_\alpha \delta^0_\beta \Bigr] \; ,
\end{eqnarray}
and the three scalar propagators, which we discuss below, obey:
\begin{equation}
\label{ScaPropDE} {\cal D}_I  \, i\Delta_I(x;x') = i\delta^D(x-x')
\;\;\;\; , \;\;\;\; I = A, B, C \; .
\end{equation}
It follows that the graviton propagator satisfies the equation,
\begin{equation}
\label{GravPropDE} {\cal D}^{\rho\sigma\mu\nu} \,
i[_{\mu\nu}\Delta_{\alpha\beta}](x;x') =
i\delta^\rho_{(\alpha}\delta^\sigma_{\beta)} \, \delta^D(x \!-\! x')
\; .
\end{equation}

To write expressions for the three scalar propagators, we note that
our gauge fixing term (\ref{gauge}) will produce a graviton
propagator that contains a de Sitter invariant part as well as a de
Sitter symmetry breaking piece. For the former it will be useful to
introduce a function $y(x;x') = 4 \sin^2[\frac12 H \ell(x;x')]$ of
the de Sitter invariant $\ell(x;x')$ between the points $x^\mu$ and
$x^{\prime\mu}$ defined by,
\begin{equation}
\label{ydef}
y(x;x') \equiv a a' H^2 \Biggl[ \Bigl\Vert \vec{x} \!-\! \vec{x'}
\Bigr\Vert^2 - \Bigl(\vert \eta \!-\! \eta'\vert \!-\! i
\epsilon\Bigr)^2 \Biggr]\; .
\end{equation}
Because $y(x;x')$ is a de Sitter invariant, any function of $y(x;x')$
is also de Sitter invariant. Moreover covariant derivatives of it are
de Sitter invariant, this means that the first three derivatives of
$y(x;x')$ produce a convenient basis of de Sitter invariant bi-tensors
\cite{KW2},
\begin{eqnarray}
\label{ybasis}
\frac{\partial y(x;x')}{\partial x^{\mu}} & = & H a \Bigl(y\delta^0_{\mu}
\!+\! 2 a' H \Delta x_{\mu} \Bigr) \; , \\
\frac{\partial y(x;x')}{\partial x'^{\nu}} & = & H a' \Bigl(y
\delta^0_{\nu}
\!-\! 2 a H \Delta x_{\nu} \Bigr) \; , \\
\frac{\partial^2 y(x;x')}{\partial x^{\mu} \partial x'^{\nu}} & = &
H^2 a a' \Bigl(y \delta^0_{\mu} \delta^0_{\nu} \!+\! 2 a' H \Delta
x_{\mu} \delta^0_{\nu} \!-\! 2 a \delta^0_{\mu} H \Delta x_{\nu}
\!-\! 2 \eta_{\mu\nu}\Bigr) \; . \qquad
\end{eqnarray}
Here and subsequently $\Delta x_{\mu} \equiv \eta_{\mu\nu}
(x \!-\!x')^{\nu}$.

It turns out that only the $A$-type propagator contains a de Sitter
breaking part as it corresponds to a massless, minimally coupled
scalar for which it is well known that no de Sitter invariant
solution exists \cite{AF}. Preserving only the symmetries of
homogeneity and isotropy this scalar propagator can be written as
\cite{OW},
\begin{equation}\label{Atype}
i\Delta_A(x;x') = A(y(x;x')) + K \ln(aa') \; ,
\end{equation}
where the constant $K$ is,
\begin{equation}\label{kc}
K \equiv \frac{ H^{D-2}}{(4 \pi)^{\frac{D}2}} \frac{\Gamma(D \!-\! 1)}{
\Gamma(\frac{D}2)} \; .
\end{equation}
The function $A(y)$ is,
\begin{eqnarray}
\label{Af}
\lefteqn{A(y) = \frac{H^{D-2}}{(4\pi)^{\frac{D}2}} \Biggl\{
\Gamma\Bigl(\frac{D}2 \!-\!1\Bigr) \Bigl(\frac{4}{y}\Bigr)^{
\frac{D}2 -1} \!+\! \frac{\Gamma(\frac{D}2 \!+\! 1)}{\frac{D}2
\!-\! 2} \Bigl(\frac{4}{y} \Bigr)^{\frac{D}2-2} \!+\! A_1 }
\nonumber \\
& & \hspace{1.3cm} - \sum_{n=1}^{\infty} \Biggl[
\frac{\Gamma(n\!+\!\frac{D}2\!+\!1)}{(n\!-\!\frac{D}2\!+\!2)
\Gamma(n\!+\! 2)} \Bigl(\frac{y}4 \Bigr)^{n - \frac{D}2 +2}
\!\!\!\!\! - \frac{\Gamma(n \!+\! D \!-\! 1)}{n \Gamma(n \!+\!
\frac{D}2)} \Bigl(\frac{y}4 \Bigr)^n \Biggr] \Biggr\} . \qquad
\label{DeltaA}.
\end{eqnarray}
Here the constant $A_1$ is defined by,
\begin{equation}
\label{cA1} A_1 = \frac{\Gamma(D\!-\!1)}{\Gamma(\frac{D}2)} \Biggl\{
- \psi\Bigl(1 \!-\! \frac{D}2\Bigr) + \psi\Bigl(\frac{D \!-\! 1}2
\Bigr) + \psi(D \!-\!1) + \psi(1) \Biggr\} .
\end{equation}
On the other hand the $B$-type and the $C$-type propagators are de
Sitter invariant,
\begin{eqnarray}
\label{Bf}
\lefteqn{i\Delta_B(x;x') \equiv B(y) = \frac{H^{D-2}}{(4\pi)^{\frac{D}{2}}}
\Gamma\left(\frac{D}{2} \!-\! 1\right) \left(\frac{4}{y}\right)^{\frac{D}{2}
\!-\!1} } \nonumber \\
& & \hspace{1.5cm} - \frac{H^{D\!-\!2}}{(4\pi)^{\frac{D}{2}}}
\displaystyle\sum_{n=0}^{\infty} \left\{
\frac{\Gamma\left(n\!+\!D\!-\!2\right)} {\Gamma\left(n \!+\!
\frac{D}{2} \right)}\left(\frac{y}{4}\right)^{n} \!-\!
\frac{\Gamma\left(n\!+\!\frac{D}{2}\right)}{\Gamma\left(n \!+\!2
\right)}\left(\frac{y}{4}\right)^{n\!-\!\frac{D}{2}+2} \right\} ,
\qquad \\
\label{Cf} \lefteqn{i\Delta_C(x;x') \equiv C(y) =
\frac{H^{D-2}}{(4\pi)^{\frac{D}{2}}}
\Gamma\left(\frac{D}{2}\!-\!1\right)
\left(\frac{4}{y}\right)^{\frac{D}{2} \!-\!1} +
\frac{H^{D\!-\!2}}{(4\pi)^{\frac{D}{2}}}
\displaystyle\sum_{n=0}^{\infty} } \nonumber \\
& & \hspace{-.5cm} \times \left\{ (n \!+\! 1)\frac{\Gamma\left(n
\!+\! D \! -\! 3\right)}{\Gamma\left(n \!+\! \frac{D}{2} \right)}
\left(\frac{y}{4}\right)^{n} \!-\! \bigg(n\!-\!\frac{D}{2} \!+\!
3\bigg)\frac{\Gamma\left(n\!+\!\frac{D}{2} \!-\! 1
\right)}{\Gamma\left(n \!+ \!2 \right)}\left(\frac{y}{4}\right)^{n
\!-\!\frac{D}{2}+2} \right\} . \qquad
\end{eqnarray}
Each of the three invariant functions $A(y), B(y)$ and $C(y)$
contains an infinite series in powers of $y$. That may seem a little
discouraging at first, but inspection reveals that each of the three
series vanishes for $D=4$. This means that we need to keep only
those terms which multiply  potentially divergent terms. Another
simplification is that our computation will require only the
coincidence limits of these functions (and their first derivatives).
To take the coincidence limit means to set $x^\mu=x^{\prime\mu}$,
hence it follows that in this limit $a=a^\prime$, $\Delta x^\mu=0$,
and $y=0$. This gives,
\begin{equation}
\label{coinlim} \lim_{x'\rightarrow x}\frac{\partial
y(x;x')}{\partial x^{\mu}} = 0 \;\; , \;\; \lim_{x' \rightarrow x}
\frac{\partial y(x;x')}{\partial x'^{\nu}} = 0 \;\; , \;\;
\lim_{x'\rightarrow x}\frac{\partial^2 y(x;x')}{\partial x^{\mu}
\partial x'^{\nu}} = -2 H^2 a^2 \eta_{\mu\nu} \; .
\end{equation}
Furthermore we recall that in dimensional regularization, any
$D-$dependent power of zero is automatically set equal to zero.
We can then summarize the coincidence limits we will need in the
following way. For the three types $(I=A, B, C)$ of propagators
$i\Delta_I(x;x')=I(y)+\delta^A_IK\ln(aa')$ the coincidence limits are,
\begin{eqnarray}
\label{cl1}
\lim_{x'\rightarrow x} i \Delta_I(x;x') & = & I(0) + \delta^A_I
\times 2 K \ln(a) \; , \\
\lim_{x'\rightarrow x} \partial_\mu i\Delta_I(x;x') & = & \delta^A_I
\times K H a \delta^0_\mu \; , \\
\lim_{x'\rightarrow x} \partial'_\nu i\Delta_I(x;x') & = & \delta^A_I
\times K H a \delta^0_\nu \; , \\
\label{cl2} \lim_{x' \rightarrow x} \partial_\mu \partial'_\nu
i\Delta_I(x;x') & = & I'(0) \times - 2 H^2 a^2 \eta_{\mu\nu} \; .
\end{eqnarray}
>From expressions (\ref{Af})-(\ref{Cf}) we obtain,
\begin{equation}
A(0) = \frac{H^{D-2}}{(4\pi)^{\frac{D}{2}}}\, A_1 \;\; , \;\; B(0) =
-\frac{H^{D\!-\!2}}{(4\pi)^{\frac{D}{2}}} \frac{\Gamma(D\!-\!2)}{
\Gamma(\frac{D}{2})} \;\; , \;\; C(0) =
\frac{H^{D\!-\!2}}{(4\pi)^{\frac{D}{2}}}
\frac{\Gamma(D\!-\!3)}{\Gamma(\frac{D}{2})} \; ,
\end{equation}
and
\begin{equation}
A'(0) = \frac{H^{D\!-\!2}}{(4\pi)^{\frac{D}{2}}} \frac{\Gamma(D)}{4
\Gamma\left(\frac{D}{2}\!+\!1\right)} \; , \; B'(0) = \frac{(D \!-\!
2)}{2 D} \, B(0) \; , \; C'(0) = \frac{(D \!-\! 3)}{D} \, C(0) \; .
\end{equation}

\section{The Hartree Approximation}\label{hartree}

To calculate the effects inflationary gravitons have on dynamical
gravitons we would need to calculate the graviton self-energy and
then solve the effective field equations at least to one-loop order.
That work is currently in progress. However, we can anticipate the
result by employing the Hartree, or mean-field, approximation
\cite{Howie,DDPT,PW,MW1,enh,LW}. The idea is that whenever we encounter
a product of graviton fields in the equations of motion we can
consider one graviton field as being the ``external'' one and
approximate the others in the same product by taking their
expectation value in the vacuum. To illustrate this, let us denote
an external graviton field by $E_{\alpha\beta}(x)$. The Hartree
approximation then consists of the following replacements,
\begin{eqnarray}\label{ex1}
\label{harex1} h_{\mu\nu} & \rightarrow & E_{\mu\nu} \;, \\
\label{harex2} h_{\mu\nu}h_{\rho\sigma} & \rightarrow & E_{\mu\nu}
\langle h_{\rho\sigma} \rangle + E_{\rho\sigma} \langle h_{\mu\nu}
\rangle \;, \\\label{ex3}
\label{harex3} h_{\mu\nu} h_{\rho\sigma} h_{\alpha\beta} &
\rightarrow & E_{\mu\nu} \langle h_{\rho\sigma} h_{\alpha\beta}
\rangle + E_{\rho\sigma} \langle h_{\mu\nu} h_{\alpha\beta} \rangle
+ E_{\alpha\beta} \langle h_{\mu\nu} h_{\rho\sigma} \rangle \; ,
\end{eqnarray}
for one, two and three gravitons respectively. Because it is important 
to understand what this approximation includes and what it does not, 
we digress to discuss the technique in the context of a simple quantum 
mechanical model. We then implement the approximation for quantum gravity.

\subsection{Hartree approximation for anharmonic oscillator}

The point of this sub-section is to explicate the meaning and
validity of the Hartree approximation in the context of a point
particle $q(t)$ whose Lagrangian is,
\begin{equation}
\label{ahoLag}
L = \frac{1}{2 }m \dot{q}^2(t) - \frac1{2} m \omega^2 q^2(t) -
\frac1{3} m \omega^2 g q^3(t) - \frac1{4} m \omega^2 g^2 q^4(t) \; .
\end{equation}
Here $g$ is a parameter with the dimensions of inverse length which
quantifies how far the model is from being a simple harmonic oscillator.
We first give a precise definition for the effective mode function
$u(t)$. We then compute the full result for $u(t)$ at one loop ($g^2$) 
order and compare this with what the Hartree approximation gives. The 
sub-section closes with a discussion of previous Hartree computations 
of the effective mode function in various quantum field theories.

Even though our model (\ref{ahoLag}) is not a harmonic oscillator 
for $g \neq 0 $, we can still form the initial position and velocity 
into the raising and lowering operators of a harmonic oscillator,
\begin{equation}
\label{ahoNorm}
a \equiv \sqrt{\frac{m \omega}{2 \hbar}} \Bigl[ q(0) + 
\frac{i \dot{q}(0)}{\omega} \Bigr] \qquad \Longrightarrow \qquad 
[a,a^\dagger] = 1 \; .
\end{equation}
Similarly, there is a Heisenberg state $\vert \Omega \rangle$ which 
is the normalized ground state of the harmonic oscillator as perceived
by the $t=0$ operators,
\begin{equation} 
a \vert \Omega \rangle = 0 = \langle \Omega \vert a^{\dagger} \qquad ,
\qquad \langle \Omega \vert \Omega \rangle = 1 \; .
\end{equation}
The effective mode function $u(t)$ is the matrix element of $q(t)$ 
between the ground state and the ($g = 0$) first excited state,
\begin{equation}
u(t) \equiv \langle \Omega \vert q(t) \, a^{\dagger} \vert 
\Omega \rangle = \langle \Omega \vert [q(t), a^{\dagger} ]
\vert \Omega \rangle \; . \label{modedef}
\end{equation}
In quantum field theory it would be the matrix element of the field
between free vacuum and the free one particle state, at whatever time
the system is released. One might also include perturbative corrections 
to the $t=0$ states which would alter the initial time dependence of 
$u(t)$ but not its late time form \cite{KOW}.

To derive an exact expression for $u(t)$ at order $g^2$ we require
the initial value solution of the Heisenberg equation of motion,
\begin{equation}
\label{ahoEom}
\ddot{q}(t) + \omega^2 q(t) + \omega^2 g q^2(t) + 
\omega^2 g^2 q^3(t) = 0 \; .
\end{equation}
We solve this equation by perturbatively expanding the solution in 
powers of the parameter $g$,
\begin{equation}
\label{ahoPert}
q(t) = q_0(t) + g q_1(t) + g^2 q_2(t) + g^3 q_3(t) + \dots
\end{equation}
Collecting powers of $g$, the zeroth order equation is,
\begin{equation}
\label{ahoEom0}
\ddot{q}_0(t) + \omega^2 q_0(t ) = 0 \; .
\end{equation}
We treat the initial value data as zeroth order so $q_0(t)$ is,
\begin{equation}
\label{ahoSol0}
q_0(t) = q(0) \cos(\omega t) + \frac{\dot{q}(0)}{\omega} \, \sin(\omega t)
= \sqrt{\frac{\hbar}{2 m \omega}} \Bigl( e^{-i\omega t} a + 
e^{i\omega t} a^\dagger \Bigr) \; .
\end{equation}
Similarly, the first order equation is,
\begin{equation}
\label{ahoEom1}
\ddot{q}_1(t) + \omega^2 q_1(t) = -\omega^2 q_0^2 \; .
\end{equation}
The solution to this equation is,
\begin{equation}
\label{ahoSol1}
q_1(t) = \int^t_0 \!\! dt' \, \frac{\sin[\omega(t \!-\! t')]}{\omega}
\times -\omega^2 q^2_0(t') = -\omega \!\! \int_0^t \!\! dt' \, 
\sin[\omega (t \!-\! t')] q_0^2(t') \; .
\end{equation}
In the same way the second order equation is,
\begin{equation}
\label{ahoEom2}
\ddot{q}_2(t) + \omega^2 q_2(t) = -\omega^2 q_0^3(t) 
- \omega^2 \Bigl[ q_0(t) q_1(t) \!+\! q_1(t) q_0(t) \Bigr] \; ,
\end{equation}
whose solution is found to be,
\begin{equation}
\label{ahoSol2}
q_2(t) = -\omega \!\! \int_0^t \!\! dt' \, \sin[\omega (t \!-\! t')]
\Bigl[ q_0^3(t') \!+\! q_0(t') q_1(t') \!+\! q_1(t') q_0(t') \Bigr] \; .
\end{equation}

The commutator of $q_0(t)$ with $a^{\dagger}$ facilitates our computation,
\begin{equation}
\label{q0comm}
[q_0(t),a^\dagger] = \sqrt{\frac{\hbar}{2 m \omega}} \, e^{-i\omega t} 
\equiv u_0(t) \; .
\end{equation}
With our perturbative solutions (\ref{ahoSol1}) and (\ref{ahoSol2}), 
this gives,
\begin{eqnarray}
\lefteqn{ [q(t), a^{\dagger}] = u_0(t) - 2 g \omega \!\! \int_0^t \!\! dt'
\, \sin[ \omega (t \!-\! t')] q_0(t') u_0(t') } \nonumber \\
& & \hspace{.5cm} -g^2 \omega \!\! \int_0^t \!\! dt' \, \sin[\omega 
(t \!-\! t')] \Biggl[ 3 q_0^2(t') u_0(t') \!+\! 2 q_1(t') u_0(t') 
\nonumber \\
& & \hspace{2cm} - 2 \omega \!\! \int_0^{t'} \!\! dt'' \, 
\sin[\omega (t' \!-\! t'')] \Bigl\{q_0(t'), q_0(t'')\Bigr\} u_0(t'') 
\Biggr] + O(g^3) \; . \qquad
\end{eqnarray}
Because the expectation value of any odd number of $q_0$'s vanishes, the
perturbative expansion of the effective mode function is,
\begin{eqnarray}
\lefteqn{ u(t) = u_0(t) -g^2 \omega \!\! \int_0^t \!\! dt' \, \sin[\omega 
(t \!-\! t')] \Biggl[ 3 \langle q_0^2(t')\rangle u_0(t') \!+\! 
2 \langle q_1(t') \rangle u_0(t') } \nonumber \\
& & \hspace{1.5cm} - 2 \omega \!\! \int_0^{t'} \!\! dt'' \, 
\sin[\omega (t' \!-\! t'')] \Bigl\langle \Bigl\{q_0(t'), q_0(t'')\Bigr\} 
\Bigr\rangle u_0(t'') \Biggr] + O(g^4) \; . \qquad \label{modeper}
\end{eqnarray}

Expression (\ref{modeper}) is the full one loop result. We can 
recognize the Hartree contribution by acting the kinetic operator,
\begin{eqnarray}
\lefteqn{ \Bigl[ \Bigl( \frac{d}{dt} \Bigr)^2 + \omega^2 \Bigr] u(t) 
= -g^2 \omega^2 \Bigl[ 3 \langle q_0^2(t) \rangle + 2 \langle q_1(t) \rangle
\Bigr] u_0(t) } \nonumber \\
& & \hspace{1.5cm} + 2 g^2 \omega^3 \!\! \int_0^t \!\! dt' \, 
\sin[\omega (t \!-\! t')] \Bigl\langle \Bigl\{q_0(t) , q_0(t') \Bigr\} 
\Bigr\rangle u_0(t') + O(g^4) \; . \qquad \label{ahoME}
\end{eqnarray}
The one loop Hartree contribution comes from the term $\langle q_0^2(t) 
\rangle$ on the first line of (\ref{ahoME}),
\begin{equation}
u_{\rm Hartree} = u_0(t) - 3 g^2 \omega \!\! \int_0^t \!\! dt' \, 
\sin[\omega (t \!-\! t') ] \langle q_0^2(t') \rangle u_0(t') + O(g^4) \; .
\end{equation}
Of course this is the contribution from the 4-point vertex. The other 
terms on the right hand side of expression (\ref{ahoME}) also have 
simple interpretations. The factor of $\langle q_1(t) \rangle$ gives
the contribution due to an order $g$ shift in the background field, 
and the integral on the last line represents the nonlocal contribution 
from two 3-point vertices. 

It is worth working out the three one loop contributions to $u(t)$ so 
that they can be compared in detail,
\begin{eqnarray}
\lefteqn{-3 g^2 \omega \!\! \int_0^t \!\!\! dt' \sin[\omega (t \!-\! t')]
\langle q_0^2(t') \rangle u_0(t') = \frac{g^2 \hbar}{m \omega} 
\Biggl\{ -\frac38 [1 \!+\! i 2 \omega t] u_0(t) \!+\! \frac38 u_0^*(t) 
\Biggr\} , \qquad } \label{hartone} \\
\lefteqn{-2 g^2 \omega \!\! \int_0^t \!\! dt' \, \sin[\omega (t \!-\! t')]
\langle q_1(t') \rangle u_0(t') } \nonumber \\
& & \hspace{1.3cm} = \frac{g^2 \hbar}{m \omega} \Biggl\{ \frac16 u_0(2t)
+ \frac14 [1 \!+\! i 2 \omega t] u_0(t) - \frac12 u_0(0) + \frac1{12} u_0^*(t) 
\Biggr\} \; , \qquad \label{vacshift} \\
\lefteqn{ 2 g^2 \omega^2 \!\! \int_0^t \!\! dt' \, \sin[\omega (t \!-\! t')]
\!\! \int_0^{t'} \!\! dt'' \, \sin[\omega (t' \!-\! t'')] \Bigl\langle
\Bigl\{ q_0(t') , q_0(t'') \Bigr\} \Bigr\rangle u_0(t'') } \nonumber \\
& & \hspace{1.3cm} = \frac{g^2 \hbar}{m \omega} \Biggl\{\frac16 u_0(2t) - 
\Bigl[\frac5{36} \!-\! \frac16 i\omega t\Bigr] u_0(t) - \frac1{12}
u_0^*(t) + \frac1{18} u_0^*(2 t) \Biggr\} \; . \qquad 
\label{nonloc}
\end{eqnarray}
Recall that the Hartree contribution is (\ref{hartone}). The only two 
things which seem to distinguish it from the vacuum shift (\ref{vacshift}) 
and the nonlocal contribution (\ref{nonloc}) are the absence of the 
nonoscillatory term $u_0(0)$ and the absence of higher harmonics ---
$u_0(2t)$ and $u_0^*(2t)$. The form of the dominant late time behavior
is $g^2\hbar/m\omega \times i \omega t u_0(t)$, and all three 
contributions possess it, with coefficients $-\frac38$, $+ \frac14$,
and $+ \frac16$, respectively.

From the preceding discussion we see that there is nothing particularly
distinctive about the one loop Hartree contribution (\ref{hartone})
to the effective mode function $u(t)$. It does predict the form of the 
dominant late time behavior --- $ig^2\hbar/m\omega \times i\omega 
t u_0(t)$ --- but not the numerical coefficient of this term. That is 
typical of what has been found in recent computations of quantum 
corrections to the effective mode function from inflationary scalars 
and gravitons \cite{BOW,Yukawa,SQED,MW1,enh,MW2,LW}. In some cases
--- such as the photon wave function in scalar QED \cite{SQED} ---
the Hartree result gives the correct numerical coefficient of the
dominant late time behavior. In other cases --- such as scalar
corrections to the scalar wave function \cite{BOW} and graviton
corrections to the fermion wave function \cite{MW1,enh,MW2} --- it
predicts the form of the dominant late time behavior but not the
correct numerical coefficient. And there are some cases --- such
as scalar corrections to the fermion wave function in Yukawa 
theory \cite{Yukawa} --- in which Hartree contribution vanishes
even though there are very significant late time corrections.

What the one loop Hartree approximation to the effective mode function
{\it always} gives is the contribution from the 4-point vertex. (That 
is why it happens to vanish for Yukawa.) Our reasons for considering 
it for graviton corrections to other gravitons are not that it dominates 
in any particular regime but rather:
\begin{itemize}
\item{It is vastly {\it easer} to compute than the nonlocal contribution
from two 3-point vertices; and}
\item{Whatever it gives is additively present in the full result.}
\end{itemize}
It therefore sets a sort of minimum level for what one expects for
the dominant late time behavior of the full result. We hope to have
the full result to compare in about a year's time. In the meanwhile,
it seems reasonable to explore the minimum late time effect which the 
Hartree approximation predicts.

\subsection{The effective field equation}\label{EffEQN}

The linearized, quantum corrected effective field equation for
gravitons is,
\begin{equation}
\label{LinFE} {\cal D}^{\mu\nu\rho\sigma} h_{\rho\sigma}(x) - \int
d^4x' [^{\mu\nu} \Sigma^{\rho\sigma}](x;x') h_{\rho\sigma}(x') = 0
\; ,
\end{equation}
where $-i[^{\mu\nu}\Sigma^{\rho\sigma}](x;x')$ is the graviton
self-energy. To obtain the approximate version of this we employ the
perturbation theory scheme. Expanding the total Lagrangian ${\cal
L}_{\text{inv}} + {\cal L}_{\text{gf}} + {\cal L}_{\text{ghost}}$ in
powers of $\kappa$ to the order we shall require, our equation of
motion follows from the functional derivative of the action with
respect to $h_{\mu\nu}(x)$,
\begin{equation}
\label{var1} \frac{\delta {\cal S}[h]}{\delta h_{\mu\nu}(x)} = {\cal
D}^{\mu\nu\rho\sigma} h_{\rho\sigma} + \kappa h^2 + \kappa^2 h^3 +
... = 0 \; ,
\end{equation}
where the last two terms represent collectively all possible terms
in the expansion that contain two and three factors of $h_{\mu\nu}$
respectively. In applying the Hartree approximation
(\ref{harex1})-(\ref{harex3}) to the equation above we consider the
single field in the first term as the external one, and we can
neglect all terms in the ${\cal O}(h^2)$ group since the expectation
value of one field is zero. The interesting effects come form the
${\cal O}(h^3)$ group. Here all terms will be of the form $hhh$,
$hh\partial h$ or $h\partial h\partial h$, which can be seen from
the structure of ${\cal L}_{\text{inv}}$ in (\ref{Linv}). Let us
give an example of the replacement used in the Hartree approximation
applied to a generic term from the second of these cases. The
replacement is,
\begin{eqnarray}
\label{ex}
\lefteqn{h_{\alpha\beta} h_{\mu\nu} \partial_\lambda h_{\rho\sigma}
\rightarrow E_{\rho\sigma,\lambda}(x) \lim_{x'\rightarrow x} i
[_{\alpha\beta}\Delta_{\mu\nu}](x;x') } \nonumber \\
& & \hspace{1cm} + E_{\mu\nu}(x) \lim_{x'\rightarrow x}
\partial'_\lambda
i [_{\alpha\beta} \Delta_{\rho\sigma}](x;x') + E_{\alpha\beta}(x)
\lim_{x'\rightarrow x} \partial'_\lambda i [_{\mu\nu}
\Delta_{\rho\sigma}](x;x') \; . \qquad
\end{eqnarray}
After applying similar substitutions on all terms in the
${\cal O}(h^3)$ group, the next step would be to substitute and
contract our expression for the graviton propagator (\ref{gravProp})
and then apply the coincidence limit using equations
(\ref{cl1})-(\ref{cl2}). More simplifications arise when we
impose the conditions of transversality and tracelessness on
physical gravitons, namely, ${E^{\mu\nu}}_{,\mu}=0$ and $E^\mu_\mu=0$.
We also consider only purely spatial gravitons $E_{00}=0=E_{0i}$.

Once all this has been done we can extract our effective field equation
perturbatively. To do this we can similarly expand the graviton field
in powers of $\kappa^2$,
\begin{eqnarray}
\label{hexp} E_{\alpha\beta}(x) = \displaystyle\sum_{n=0}^\infty
\kappa^{2n} E^{(n)}_{\alpha\beta}(x) \; .
\end{eqnarray}
Substituting the expression for the kinetic operator (\ref{kinop}) in
(\ref{var1}) and collecting powers of $\kappa^2$ give us the zeroth
order equation,
\begin{eqnarray}\label{order0}
{\cal D}_A E^{(0)}_{ij}(x) = 0 \; .
\end{eqnarray}
Similarly the order $\kappa^2$ equation is,
\begin{eqnarray}\label{order1}
\frac12{\cal D}_A E^{(1)}_{ij}(x) + a^{D-2} {\cal D}_\eta
E^{(0)}_{ij}(x) = 0 \; .
\end{eqnarray}
The operator ${\cal D}_{\eta}$ has the form,
\begin{eqnarray}
\label{op}
\lefteqn{{\cal D}_\eta = \left(c_1 + \alpha_1 \ln a\right)
\partial^2_\eta + \left(c_2 + \alpha_2\ln a\right) (D\!-\!2) H a
\partial_\eta } \nonumber \\
& & \hspace{6cm} -\left(c_3 + \alpha_3 \ln a \right)
\partial^l \partial_l + c_4 H^2 a^2 \; , \qquad
\end{eqnarray}
where $c_i$ and $\alpha_i$ are constants whose expressions we omit
here in favor of writing below only those terms that will contribute
the most in the late-time limit. We will give solutions to these
equations in the next section.

\section{The One Loop Mode Function}\label{Deltau}

This section comprises our main result. Here we present and solve
the one-loop order graviton mode function equation in the late-time
regime. To solve this equation we will consider a spatial plane-wave
expansion for the graviton field in terms of its mode function
$u(\eta,k)$ and the same transverse, traceless and purely spatial
polarization tensor $\epsilon_{\alpha\beta}$ as in flat space,
\begin{eqnarray}
\label{plane} E_{\alpha\beta}(x) = \epsilon_{\alpha\beta}
u(\eta,k)e^{i{\vec k} \cdot {\vec x}} \; .
\end{eqnarray}
The mode functions $u(\eta,k)$ have a similar perturbative expansion,
\begin{eqnarray}
\label{uexp} u(\eta,k) = \displaystyle\sum_{n=0}^\infty \kappa^{2n}
u^{(n)}(\eta,k) \; .
\end{eqnarray}
Substituting this expansion in (\ref{plane}) and expanding the
operator ${\cal D}_A$ according to its definition (\ref{DA}),
the zeroth order equation (\ref{order0}) becomes,
\begin{eqnarray}
\label{mode0eqn} \left[\partial^2_\eta + (D\!-\!2) H a
\partial_\eta + k^2 \right] u^{(0)}(\eta,k) = 0 \; .
\end{eqnarray}
The solution to this mode equation is well known in terms of Hankel
functions $H^{(1)}_\nu(z)$,
\begin{eqnarray}
\label{mode0sol} u^{(0)}(\eta,k) = \sqrt{\frac{\pi}{4H}}
a^{-\frac{D-1}{2}} H^{(1)}_{\frac{D-1}{2}} \bigg(\frac{k}{Ha} \bigg)
\; .
\end{eqnarray}
We may as well specialize $u^{(0)}(\eta,k)$ to $D=4$, and its late
time behavior is of crucial importance for us,
\begin{eqnarray}
D = 4 \quad \Longrightarrow \quad u^{(0)}(\eta,k) & = &
\frac{H}{\sqrt{2 k^3}} \Bigl(1 - \frac{ik}{H a} \Bigr)
\exp\Bigl[\frac{ik}{H a} \Bigr] \; , \label{D=4mode} \\
& = & \frac{H}{\sqrt{2 k^3}} \Biggl[ 1 + \frac12 \Bigl(
\frac{k}{Ha}\Bigr)^2 + \frac{i}{3} \Bigl( \frac{k}{Ha}\Bigr)^3 +
\dots \Biggr] \; . \qquad \label{latemode}
\end{eqnarray}

The one-loop order mode equation follows similarly. Collecting terms
of order $\kappa^2$ we can express the one loop corrections in terms
of a differential operator $\mathcal{D}_{\eta}$ acting on
$u^{(0)}(\eta,k)$,
\begin{eqnarray}
\label{mode1eqn} \left[\partial^2_\eta + (D\!-\!2) H a
\partial_\eta + k^2 \right] u^{(1)}(\eta,k) - 2 {\cal D}_\eta
u^{(0)}(\eta,k) = 0 \; .
\end{eqnarray}
Because we are interested in the late-time limit of this equation we
need consider only the most relevant terms in ${\cal D}_\eta
u^{(0)}$, which are those that exhibit the largest growth with time
during this period. From (\ref{op}), the leading contribution comes
from the $c_4H^2a^2$ term. However, the fact that this term survives
for $k=0$ means that it must be removed by the same counterterm
which completely absorbs the one loop 1-point function in the same
gauge \cite{1p}. Hence the $c_4 H^2 a^2$ term does not contribute at
all after renormalization.

The next leading terms are those proportional to $\ln a$. Keeping
only the late-time relevant terms, the operator ${\cal D}_\eta$ is,
\begin{equation}
\label{lateop} {\cal D}_\eta = \frac{(D^3 \!-\! 12 D^2 \!+\! 31 D
\!-\! 4)}{8(D \!-\!3)} \times 2 K \ln a \times \big[
\partial^2_\eta + (D\!-\!2) H a
\partial_\eta \big] \; .
\end{equation}
Here we have neglected terms proportional to $(D-4)$ and we remind
the reader that the constant $K$ was defined in expression
(\ref{kc}). Hence the late time form of our first order mode
equation (\ref{mode1eqn}) becomes
\begin{eqnarray}
\label{mode1eqn2}
\lefteqn{ \left[ \partial^2_\eta + (D\!-\!2) H a \partial_\eta + k^2
\right] u^{(1)}(\eta,k)} \nonumber \\
& & = \frac{(D^3-12D^2+31D-4)}{2(D-3)} K \ln a \times \big[
\partial^2_\eta + (D\!-\!2) H a \partial_\eta \big] u^{(0)}(\eta,k)
\;  \qquad \\
\label{mode1eqn3} & & = -\frac{(D^3 \!-\!12 D^2 \!+\! 31 D \!-\!
4)}{2 (D\!-\!3)} K \ln a \times k^2 u^{(0)}(\eta,k) \; .
\end{eqnarray}
The last equality follows upon substitution of the zeroth order
mode equation (\ref{mode0eqn}).

At this point we note that there are no divergent terms when
$D\rightarrow4$. As a consequence we can specialize equation
(\ref{mode1eqn3}) to $D=4$ to obtain,
\begin{eqnarray}
\label{mode1eqnD4} \left[\partial^2_\eta + 2 H a
\partial_\eta + k^2 \right] u^{(1)}(\eta,k) = \frac{H^2}{2\pi^2} \times
\ln a \times k^2 u^{(0)}(\eta,k) \; . \label{u1eqn}
\end{eqnarray}
To obtain the leading late time behavior of $u^{(1)}(\eta,k)$ recall
from (\ref{latemode}) that $u^{(0)}(\eta,k)$ approaches the constant
$H/\sqrt{2 k^3}$ at late times. Hence the right hand side of
(\ref{mode1eqnD4}) grows like $\ln(a)$ at late times. Now consider
acting the differential operator on the left hand side on
$\ln(a)/a^2$,
\begin{equation}
\left[\partial^2_\eta + 2 H a
\partial_\eta + k^2 \right] \left(\frac{\ln a}{a^2} \right) = - 2 H^2
\ln a - H^2 + k^2 \frac{\ln a}{a^2} \; .
\end{equation}
The last two terms can be neglected when compared to the leading term
$\ln a$. Hence up to first order, the leading late-time limit
contribution of the graviton mode function can be written as
\begin{equation}
\label{result} u(\eta,k) = u^{(0)}(\eta,k) + \kappa^2
u^{(1)}(\eta,k) = \bigg(1 - \frac{4k^2}{\pi H^2} \times
\frac{GH^2\ln a}{a^2} \bigg)u^{(0)}(\eta,k) \; .
\end{equation}
This result gives us a rough idea about the enhancement inflationary
gravitons acquire in the late-time regime from other gravitons. We
expect that the same time dependence will be present in the full result,
the only discrepancy being a different numerical factor. Only the fully
renormalized and dimensionally regulated calculation will tell us how
good the approximation really is.

\section{Epilogue}\label{Epilogue}

We have used perturbative quantum gravity on de Sitter background
to calculate the effects of inflationary gravitons on dynamical
gravitons at one loop order in the late-time regime. We decomposed
the graviton field using a plane wave expansion and we employed the
Hartree approximation \cite{Howie,DDPT,PW,MW1,enh,LW} to obtain the
first order graviton mode equation (\ref{mode1eqnD4}). In equation
(\ref{result}) we have found that the time dependence of the
graviton mode functions is modified by a factor of $GH^2 \ln(a)/a^2$
which decays exponentially with time.

Our result (\ref{result}) might be thought to demonstrate the
irrelevance of quantum loops corrections to gravitons, but that is
not so. It {\it does} show that there are no significant loop
corrections to the tensor power spectrum, if we define
$\Delta^2_{h}(k)$ using the norm-squared of the graviton mode
function \cite{MP}, because the late time limit of the norm-squared
of the mode function is identical to its tree order result,
\begin{equation}
\Delta^2_{h}(k) = \frac{k^3}{2 \pi^2} \times 64 \pi G \times
\Bigl\vert u(\eta,k) \Bigr\vert^2_{\eta \rightarrow 0} =
\frac{16}{\pi} \, G H^2 \; .
\end{equation}
However, let us instead compare the curvature of the zeroth order
term with the curvature induced by the quantum correction. For
transverse-traceless graviton fields $h_{\mu\nu}(\eta,\vec{x})$ the
linearized Weyl tensor is,
\begin{equation}
C^{\rm lin}_{\rho\sigma\mu\nu} = -\frac{\kappa}{2 a^2} \Bigl(
h_{\rho \mu , \sigma \nu} \!-\! h_{\mu \sigma , \nu \rho} \!+\!
h_{\sigma \nu , \rho \mu} \!-\! h_{\nu \rho , \mu \sigma} \Bigr) \;
.
\end{equation}
Now specialize to a spatial plane wave $h_{ij}(\eta,\vec{x}) =
\epsilon_{ij} u(\eta,k) e^{i \vec{k} \cdot \vec{x}}$ with purely
spatial polarization, and examine the ``electric'' components,
\begin{equation}
C^{\rm lin}_{0 i 0 j} = -\frac{\kappa}{2 a^2} \Bigl(
\partial_{\eta}^2 h_{ij} \Bigr) \; . \label{electric}
\end{equation}
Next use relations (\ref{latemode}) and (\ref{result}) to compare
the late time limits of second (conformal) time derivatives of the
0th and 1st order mode functions,
\begin{eqnarray}
\partial_{\eta}^2 u^{(0)}(\eta,k) & \longrightarrow &
\frac{H k^2}{\sqrt{2 k^3}} \times 1  \; , \label{d2u0} \\
\partial_{\eta}^2 u^{(1)}(\eta,k) & \longrightarrow &
\frac{H k^2}{\sqrt{2 k^3}} \times -\frac{4}{\pi} \, G H^2 \Bigl[2
\ln(a) \!-\! 3\Bigr] \; . \qquad \label{d2u1}
\end{eqnarray}
By substituting (\ref{d2u0}-\ref{d2u1}) into the electric components (\ref{electric}) we see two things:
\begin{itemize}
\item{That the magnitude of the one loop corrections to the electric
components of the linearized curvature grows (without bound)
relative to the tree order result; and}
\item{That the one loop correction tends to cancel the tree order
result.}
\end{itemize}
So it seems fair to conclude that quantum corrections make
geometrically significant changes to gravitons. This might be
important in trying to understand how two loop effects --- which
include the gravity sourced by these one loop perturbations ---
might induce a secular change in the expansion rate
\cite{TW4}.\footnote{It might be argued that the prefactor of
$1/a^2$ in expression (\ref{electric}) makes $\ln(a)$ corrections
irrelevant as a source of corrections to the de Sitter background.
However, one must recall that (\ref{electric}) represents the effect
from {\it a single} graviton. The actual source comes from adding up
the contribution from all super-horizon gravitons, and this sum
compensates the factor of $1/a^2$, to leave the $\ln(a)$.}

It is interesting to compare what we have found about how
inflationary gravitons affect other gravitons (in the one loop
Hartree approximation) with how they affect other particles. Recall
that we found a fractional correction to the mode function of the
form $-G H^2 \ln(a)/a^2$. The effects of gravitons on massless
fermions produce a fractional secular growth in the field strength
of the form $+G H^2 \ln(a)$ \cite{MW1,enh}. By contrast, the fractional
change (in the Hartree approximation) on the photon wave function is
of the form $-G H^2 \ln(a)/a$ \cite{LW}. Just like what we found for
gravitons, the one loop correction to  photon mode functions goes to
zero, but the effect on the electric components of the relevant
field strength grow in magnitude \cite{LW}. In each of these three
cases the particles being followed have spin, which seems to be why
they continue to interact with inflationary gravitons even when
their kinetic energies have red-shifted to zero \cite{MW2}.

To conclude this work we would like to comment on the current
situation concerning the de Sitter (non-)invariance of the graviton
propagator. It is worthwhile to note that the $-GH^2\ln (a)/a^2$
enhancement we have found for dynamical gravitons arises from the
fact that, as can be seen from eqn. (\ref{mode1eqnD4}), the first
order correction to the mode functions is sourced by a factor of
$\ln a$ in the late time limit. This can be traced back to the de
Sitter breaking logarithmic term in the $A$-type scalar propagator
(\ref{Atype}). We will summarize the long controversy
\cite{HMM,MTW4} about the existence of such symmetry breaking terms
in the graviton propagator.

Mathematical physicists have for decades believed that the graviton
propagator must be de Sitter invariant because they could use
analytic continuation techniques to find explicit, de Sitter
invariant solutions for it when they add de Sitter invariant gauge
fixing terms to the action \cite{math}. Researchers who approach the
problem from the perspective of cosmology have been equally
convinced that there must be de Sitter breaking because free
dynamical gravitons obey the same equation as massless, minimally
coupled scalars \cite{Grishchuk}, which possesses no normalizable,
de Sitter invariant states \cite{AF}. Indeed, the $A$-type
propagator is precisely that of a massless, minimally coupled scalar
in the homogeneous and isotropic state required by cosmology
\cite{OW}, and its presence in any valid graviton propagator is
required by the scale invariance of the tensor power spectrum
\cite{MTW4}. Although our particular graviton propagator was derived
in a de Sitter breaking gauge \cite{TW2}, one can show that its de
Sitter breaking is physical by adding the compensating gauge
transformation \cite{Kleppe}.

The two views have been converging recently because it has been
demonstrated that there is an obstacle to adding invariant gauge
fixing terms on any manifold which possesses a linearization
instability such as de Sitter \cite{MTW1}. Ignoring the problem in
scalar quantum electrodynamics leads to unphysical, on-shell
singularities for one loop scalar self-mass-squared \cite{KW2} and
would produce similar problems in quantum gravity. It has also been
shown that the analytic continuation techniques employed by
mathematical physicists automatically subtract power law infrared
divergences to produce formal solutions to the propagator equation
which are not true propagators in the sense of being the expectation
value, in some normalized state, of the time-ordered product of two
field operators \cite{MTW2}.

It is still valid to employ de Sitter invariant gauge conditions
which are ``exact''; that is, the condition is enforced as a strong
operator equation. When this was done, without using invalid
analytic continuations, the result was a de Sitter breaking
propagator \cite{MTW3}, whose spin two part agrees with the one we
used \cite{KMW}. The same result persists for the entire 1-parameter
family of exact, de Sitter invariant gauges \cite{generalprop}.

Mathematical physicists have conceded the point about gauge fixing,
but some of them still insist on the validity of analytic
continuation because it does produce solutions to the propagator
equation \cite{Higuchi}. A recent paper by Morrison \cite{morrison}
has identified precisely the two deviations which would convert the
cosmological derivation of a de Sitter breaking propagator
\cite{MTW3,generalprop} into the derivation of a de Sitter invariant
result. One of these deviations corresponds to regarding the scalar
propagator for any $M^2$ as a de Sitter invariant and meromorphic
function of $M^2$, even for the tachyonic case of $M^2 < 0$
\cite{perils}. The other deviation corresponds to adding a constant
to the scalar equation for the spin two structure function, when no
such constant can be added for any other slow roll parameter
$\epsilon(t) \equiv -\dot{H}/H^2$ \cite{perils}. Thus we feel
confident in adopting the de Sitter breaking propagator.

\vskip 1cm

\centerline{\bf Acknowledgements}

This work was partially supported by European Union program Thalis
ESF/NSRF 2007-2013 MIS-375734, by European Union (European Social
Fund, ESF) and Greek national funds through the Operational Program
``Education and Lifelong Learning'' of the National Strategic Reference
Framework (NSRF) under ``Funding of proposals that have received a
positive evaluation in the 3rd and 4th Call of the ERC Grant Schemes'',
by NSF grant PHY-1205591, and by the Institute for Fundamental Theory
at the University of Florida.

\end{document}